# The influence of antiferromagnetism, soft out-of plane phonons and heavy electrons on the superconducting pairing mechanism of $Ba_{1-x}K_xFe_2As_2$

Chi Ho Wong* & Rolf Lortz*

Department of Physics, Hong Kong University of Science and Technology, Hong Kong, China

*roywch@gmail.com, *lortz@ust.hk

**Abstract**

Based on ab-initio calculated parameters, we apply a theoretical model on the iron-based $BaFe_2As_2$ superconductor that takes into account dramatic enhancements of the electron-phonon coupling of soft transverse phonons in the FeAs layers and antiferromagnetism. Our model is able to reproduce the $T_c$ values of $BaFe_2As_2$ found under pressure in experiments. To calculate the $T_c$ of the K-doped $Ba_{1-x}K_xFe_2As_2$ system as a function of the K content, we additionally consider the experimentally observed effective mass enhancements and Kondo temperatures in the strongly over-doped region ($0.8 < x < 1$), which decouple the antiferromagnetism and electron-phonon scattering. The highest theoretical $T_c$ at the optimal doping concentration is reproduced after optimization of antiferromagnetic fluctuations and electron-phonon coupling. Our model is also able to reproduce the dip-like structure in $T_c$ in the region where a re-entrant tetragonal phase of $C_4$ symmetry is found ($0.24 < x < 0.28$) and indicates the weakening effect of local exchange correlation energy as responsible for $T_c$ reduction. Our model thus demonstrates that the high transition temperatures and the exact doping and pressure dependence of this iron-based superconductor can be explained within an extended electron-phonon coupling model in which the structural, magnetic and electronic degrees of freedom are strongly intertwined.

**Introduction**

$AFe_2As_2$, wherein A can be Ba, Sr or Ca, is the parent compound of the '122' family of iron-based superconductors. $BaFe_2As_2$ (Ba122), which is at the focus of this article, is non-superconducting at ambient pressure with a so-called stripe-type antiferromagnetic spin density wave (SDW) order. It can be driven superconducting by application of hydrostatic external pressure, chemical internal pressure (e.g. by isovalent doping when Fe is substituted by Ru [1] or As by P [2]), or by ionic substitution in the systems $Ba_{1-x}K_xFe_2As_2$, $Ba_{1-x}Na_xFe_2As_2$ or $Ba(Fe_{1-x}Co_x)_2As_2$. Substitution of $Ba^{2+}$ by $K^+$ or $Na^+$ introduces holes [3]. The substitution of $Fe^{2+}$ by $Co^{3+}$ [4] introduces electrons. Upon application of pressure or doping, the SDW transition is gradually suppressed and a superconducting phase develops around the point where this transition is extrapolated to zero temperature. The magnetic transition almost coincides with a structural transition from a tetragonal high-temperature to an orthorhombic low-temperature structure [5] identified as a nematic electronic transition in which electrons spontaneously reduce the four-fold rotational symmetry in the FeAs plane to a two-fold symmetry. The phase diagram of hole-doped Ba122 is particularly rich with several additional phases occurring near the point where the SDW transition meets the superconducting phase boundary. For example, in $Ba_{1-x}K_xFe_2As_2$ a re-entrant tetragonal $C_4$ phase region appears in which the high-temperature four-fold symmetry in the plane is restored, presumably driven by a re-arrangement of the electron spins and the electronic nematic order [6,7]. The presence of this phase somewhat suppresses the superconducting transition [7,8]. Superconductivity thus emerges in the vicinity of a rich variety of other electronic ordered phases, namely the antiferromagnetic SDW and the nematic phases in $Ba_{1-x}K_xFe_2As_2$, which is a hallmark of unconventional superconductivity with a rich interplay of structural, magnetic and electronic degrees of freedom [9].

Although unconventional superconductivity is generally regarded to be of a non-phonon-mediated origin, recent work suggests that the role of electron-phonon coupling may has been underestimated: B. Li *et al.* showed that the phonon softening of AFeAs (A: Li or Na) under antiferromagnetic background causes an increase of the electron-phonon coupling by a factor of ~2 [10]. S. Deng *et al.* [10] treated the out-of-plane lattice vibration as a phonon softening phenomenon in order to amplify electron-phonon scattering in their first-principle linear response calculation. Coh *et al.* refined these models further [10,11] and suggested that the electron phonon coupling of iron-based superconductors was underestimated by a factor of 4. They claimed that the cancellation of the electron-phonon coupling with the nearest neighbors [12] under antiferromagnetic background leads to a first amplification factor of 2, and the abnormal out-of-plane vibration of the iron atom actuated by tetrahedral sites increases the electrical potential by a further factor of 100/45 ~ 2 [12]. However, neither the role of the out-of-plane phonon nor the model of a magnetically-assisted electron-phonon coupling has received much attention from the research community, since the reported theoretical $T_c$ values were far below the experimental values [10,11].

Recently, we reported an even more refined model combining an antiferromagnetically assisted electron-phonon coupling [13] with the effect of the soft out-of-plane phonon triggered by tetrahedral sites [11,12]. It provides a pairing strength formula [13] that has made it possible to derive theoretical $T_c$ values for NaFeAs, LiFeAs and FeSe corresponding to those observed experimentally. We considered that the electron-phonon scattering matrix, due to the high transition temperatures, must take into account the full electronic DOS in a range of $E_F - E_{Debye}$ to $E_F$ and not only the Fermi level value. $E_{Debye}$ represents the upper limit of the phonon energies that can be transferred to electrons, and at the high transition temperatures of Fe-based superconductors, contributions of high-energy phonons in the electron-phonon scattering mechanism become important in contrast to classical low-$T_c$ superconductors. This approach, which is a direct consequence energy conservation, is supported by experiments: A shift of the spectral weight between the normal and the superconducting state is clearly visible in the photoemission spectra below the superconducting energy gap of various iron-based compounds in an energy range of ~30 - 60 meV below the Fermi energy [14-17]. This energy range is approximately in the order of the Debye energy. This made it possible to simplify the complicated magnetic interactions as a function of pressure in NaFeAs, LiFeAs and FeSe to the solution of the Ising Hamiltonian with the negative sign of the antiferromagnetic exchange coupling [13]. In order to investigate whether our approach can describe more complex electronic phenomena in other iron-based superconductors, the $Ba_{1-x}K_xFe_2As_2$ (122-type) with the highest $T_c$ among the 122 type iron-based superconductors [8] with its rich phase diagram is investigated here. The space group of $BaFe_2As_2$ is changed from I4/mmm to F/mmm above 0.8GPa and the $BaFe_2As_2$ in the F/mmm state becomes superconducting [18]. The experimental Mössbauer spectrum confirmed the existence of a double-$Q$ magnetic structure in the Na-doped $AFe_2As_2$ series (A: Ba, Sr) [6,7]. The double-$Q$ magnetic order and a chequerboard charge order in the re-entrant tetragonal phase were also observed in $Ba_{0.73}K_{0.24}Fe_2As_2$ [7,8], where it lead to a significant reduction of $T_c$. Furthermore, a Kondo effect with significant effective mass enhancement must be considered in the heavily doped $Ba_{1-x}K_xFe_2As_2$: In the end member $KFe_2As_2$ the effective mass enhancement [19,20] with a Sommerfeld coefficient increased to over 100 mJ·mol·$K^2$ and a Kondo temperature of 165K was reported [19-21]. Both the magnetic and thermodynamic experiments on $KFe_2As_2$ showed a heavy fermion behavior, which leads to a massive reduction in the local magnetic moment.

This article deals with three tasks. First, we will examine whether the pairing strength formula [13] can yield the theoretical $T_c$ values of $Ba_{1-x}K_xFe_2As_2$ as a function of $x$ and as a function of external hydrostatic pressure at reasonable values. We will refine the $T_c$ calculation in the $C_4$ magnetic phase by monitoring the local exchange coupling when the local Fe moments point to the out-of-plane direction [7]. Finally, we test whether the low $T_c$ values in the highly doped region are due to a decoupling between the electron-phonon scattering and antiferromagnetism.

**Computational algorithm**

Our recently reported algorithm of iron-based superconductivity is reviewed here [13]. As elaborated from B. Li *et al.* and S. Deng *et al.*, Coh, Cohen and Louie have proposed that the antiferromagnetic background in the presence of tetrahedral sites increases the electron-phonon coupling of iron-based superconductors by a factor of 4, hereinafter referred to as Coh factor $C_F$ [12]. However, using the Coh factor itself is not sufficient to obtain reasonable theoretical $T_c$ values of $Ba_{1-x}K_xFe_2As_2$ unless the electron-phonon interaction is not limited to the Fermi level [13].

According to the ARPES data [14-17], numerous iron-based superconductors, including $Ba_{1-x}K_xFe_2As_2$, share a common feature, namely an obvious shift of the spectral weight in the photoemission spectra, which is visible in an energy range down to 30 - 60 meV below the Fermi energy [15,16]. Therefore, we consider electron-phonon scattering in multi-energy layers. The amplified electron-phonon coupling is described as $\lambda_{PS} = 2\int \alpha_{PS}^2 \frac{F(\omega)}{\omega} d\omega$ where $\alpha_{PS} = \alpha_{E_F} C_F R_g$ [13]. The $R_g$ factor controls the amount of electrons scattered below the Fermi level $E_F$ [13]. If it is a phonon-mediated superconductor, the maximum energy to activate the electrons below $E_F$ cannot exceed the Debye energy $E_{Debye}$. We define $R_g = \dfrac{\left\langle \sum_{-\infty}^{E_F} g(E)\delta_A(E) \right\rangle}{\left\langle \sum_{-\infty}^{E_F} g(E)\delta_B(E) \right\rangle}$ where $\delta_A(E) = 1$ if $(E_F - E_{Debye}) \geq E \geq E_F$. Similarly, $\delta_B(E) = 1$ if $E = E_F$. Otherwise $\delta_A(E) = \delta_B(E) = 0$. The electron phonon scattering term on the Fermi surface is labeled as $\alpha_{E_F}$ with the Coh factor $C_F = 4$ [12] (unless otherwise stated). In the case of a strong coupling, the electron-phonon coupling and the Coulomb pseudopotential $\mu$ are renormalized by $\lambda_{PS} + 1$ [13]. We define $M_{Fe}$ and $E_{co}$ as the magnetic moment of the Fe atom and the exchange-correlation energy, respectively. Thus the pairing strength formula of Ba$_{1-x}$K$_x$Fe$_2$As$_2$ becomes [13]:

$$\lambda = {}^*\lambda_{PS} f(E_{ex}) \quad \text{where} \quad f(E_{ex}) \sim \frac{[M_{Fe} M_{Fe} E_{co}]_{P>0}}{[M_{Fe} M_{Fe} E_{co}]_{P=0}}.$$

The local Fe moments point to the out-of-plane direction in the doping range of $0.24 < x < 0.29$ [7]. This should change the local exchange-correlation energy. We define the exchange-correlation energy as $E_{co}(a,b,c)$, where $ab$ refers to the unit lengths within the $xy$ FeAs plane and $c$ is the layer to layer distance. The axis designation is illustrated in Figure 1. In ordinary iron-based superconductors, the ferromagnetic Fe chain with the head-to-tail orientation of the spins (renamed 'HT' Fe chain) couples antiferromagnetically to the other Fe chains along the $b$-axis [22-26]. To calculate the $T_c$ in the range $0.24 < x < 0.29$, we estimate a trial pairing strength at which the head-to-tail alignment of spins is virtually directed along the $b$-axis. Then the head-to-tail alignment of spins is reset to the $c$-axis under the same lattice constant and then the loss of local exchange coupling becomes comparable. The trial pairing strength is corrected by the loss of exchange coupling. To achieve this, the isolated 'HT' Fe chain is treated as an energy source to activate the antiferromagnetic coupling. We define the maximum exchange-correlation energy of an isolated 'HT' Fe chain as $E_{max} = |E_{co}(\infty,b,\infty)|$. When the 'HT' Fe chains are coupled, the $E_{max}$ is partly transferred to the antiferromagnetic exchange coupling $E_{AF}(a) = n_{co}^a E_{max}$ and

$E_{AF}(c) = n_{co}^c E_{max}$. The fractions $n_{co}^a$, $n_{co}^b$ and $n_{co}^c$ are the coefficients of the energy-transfer function with the relation of $[n_{co}^a, n_{co}^b, n_{co}^c] \propto [e^{-Ax}, e^{-By}, e^{-Cz}]$. The decay rates along the x-, y- and z-direction are controlled by the real constants A, B and C. The ferromagnetic exchange coupling, meanwhile, is expressed as $E_{ferro}(b) = n_{co}^b E_{max}$. The energy partitions are expressed as $E_{max} = E_{AF}(a) + E_{ferro}(b) + E_{AF}(c)$.

For example, the lattice constants of $Ba_{0.75}K_{0.25}Fe_2As_2$ are imported as $a = 3.93$Å, $b = 3.93$Å and $c = 13.22$Å. The strength of the trial pairing is registered by aligning the Fe moments along the b axis in the 'HT' format. We estimate the trial exchange-correlation energy, where $E_{max}^{trial} = E_{AF}^{trial}(a) + E_{ferro}^{trial}(b) + E_{AF}^{trial}(c)$. Next, we correct the pairing strength by resetting the spin configuration to the experimentally observed "side by side" format along the b-axis and recalculate the exchange-correlation energy. This makes the ratios of the exchange-correlation energy become acquirable, i.e. $\frac{E_{AF}(a)}{E_{AF}^{trial}(a)} \sim \frac{n_{co}^a E_{max}}{n_{co}^{a'} E_{max}^{trial}}$, $\frac{E_{ferro}(b)}{E_{ferro}^{trial}(b)} \sim \frac{n_{co}^b E_{max}}{n_{co}^{b'} E_{max}^{trial}}$ and $\frac{E_{AF}(c)}{E_{AF}^{trial}(c)} \sim \frac{n_{co}^c E_{max}}{n_{co}^{c'} E_{max}^{trial}}$.

To find $E_{max}$ and $E_{max}^{trial}$, we isolate the Fe chains completely so that we consider them as a 1D chain, where the spin configuration must be in the 'HT' format to favor the lowest magnetic energy. To allow a fair comparison, the maximum exchange couplings are important in 'HT' format as $E_{max}^{trial} = |E_{co}(\infty, 0.278nm, \infty)|$ and $E_{max} = |E_{co}(\infty, \infty, 1.322nm)|$.

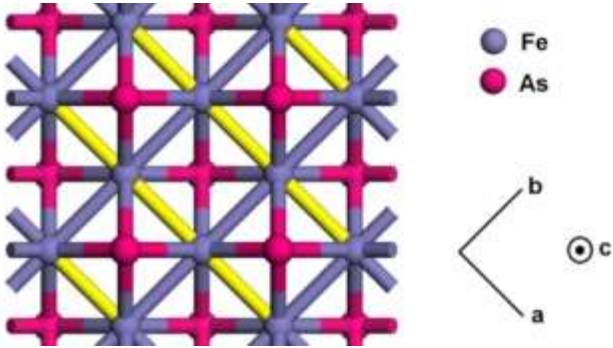

**Figure 1.** The local area of the FeAs layer. The yellow bars refer to the antiferromagnetic coupling. The heavy fermion factor occurs at $x > 0.73$ with a Kondo temperature of ~10K [21]. In this case, the antiferromagnetic enhancement of the electron-phonon coupling is suppressed and the Coh factor will be reduced to 2, the only remaining enhancement being the soft out-of-plane phonon. If the electron-phonon coupling and the antiferromagnetism are decoupled, the $\lambda$ correspond to $^*\lambda_{PS}$.

The Debye temperature is calculated by $T_{Debye} = \frac{h}{2\pi k_B}\left[\frac{18\pi^2}{V}\frac{1}{\sum(1/v_s^3)}\right]^{1/3}$ where $h$, $k_B$, $V$, $v_s$ are the Planck constant, Boltzmann constant, the unit cell volume and the speed of sound, respectively [27]. The DFT calculation is done with the WIEN2K package [28-30]. The electronic properties are compiled by the GGA-PBE functional and the finite displacement method is used to calculate the phonon dispersion. The pairing strength $\lambda$ is substituted into the McMillian $T_c$ formula [31]. All Coulomb pseudopotential are set to 0.15 because the highly correlated electron-electron interaction may not be easily calculated from the electronic DOS, Debye temperature and Fermi energy [32]. However, the consideration of pseudopotentials in the range of 0.1 to 0.2 should make sense, with a maximum error affecting our theoretical $T_c$ calculation of not more than ~15% [13].

**Results**

Fig. 2a shows our theoretical $T_c$ of BaFe$_2$As$_2$ under external pressure in comparison to experimental literature data [18]. The enhanced electron-phonon coupling increases slightly with increasing pressure. The exchange Hamiltonian is optimized at 1.3GPa, but is drastically reduced at high pressure, as shown in Fig. 2b. The calculated Debye temperatures of uncompressed BaFe$_2$As$_2$ in the limit of low and high temperatures are 379K and 470K, respectively [33]. Despite our computed $\lambda_{E_F}$ of BaFe$_2$As$_2$ at 0.8GPa is 0.33, the Coh and $R_g$ factors increase the pairing strength to 0.779 to allow the theoretical $T_c$ to occur at 29K.

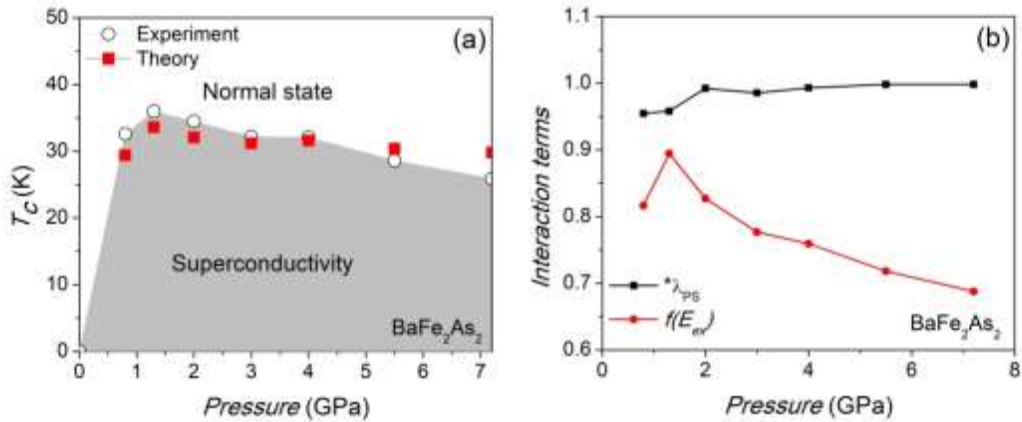

**Figure 2.** The $T_c$ distribution and the corresponding interaction terms of BaFe$_2$As$_2$ as a function of hydrostatic pressure. (a) The theoretical $T_c$ of BaFe$_2$As$_2$ under compression (squares) along with experimental data (open circles) [18]. (b) The effect of pressure on the interaction terms. The Debye temperatures at 300K are used.

The experimental and theoretical $T_c$ values of Ba$_{1-x}$K$_x$Fe$_2$As$_2$ agree well, as shown in Figure 3 [8]. We import the entire set of lattice parameters measured by Rotter *et al.* as a function of doping [26]. Although the $T_c$ values at $0.2 < x < 0.3$ are missing in the Rotter *et al.* article, we capture the same $T_c$ shift in percent at $0.23 < x < 0.29$ reported by Böhmer *et al.* [8] to allow a more systematic comparison in $T_c$. According to our model, the dip-like structure in $T_c$ at $0.23 < x < 0.29$ is producible after we compute $\left.\frac{E_{AF}(a)}{E_{AF}^{trial}(a)}\right|_{x=0.25} = 0.82$ and $\left.\frac{E_{AF}(a)}{E_{AF}^{trial}(a)}\right|_{x=0.27} = 0.85$. The Ba$_{0.7}$K$_{0.3}$Fe$_2$As$_2$ and Ba$_{0.6}$K$_{0.4}$Fe$_2$As$_2$ show a theoretical $T_c$ of 37.3K and 36.8K, respectively. Figure 3b illustrates the individual components of the pairing strength. The $^*\lambda_{PS}$ reaches a maximum at $x \sim 0.25$ [34]. When the doping concentration is increased from 0.25 to 0.6, the $^*\lambda_{PS}$ drops slightly from 0.993 to 0.990, but a rapid decrease in $^*\lambda_{PS}$ can be observed at $x > 0.6$. On the other hand, the $f(E_{ex})$ is not linearly increased from $x = 0$ to $x = 0.6$ where an unexpected drop in $f(E_{ex})$ is observed at $0.24 < x < 0.28$. A decrease in the theoretical $T_c$ is observed in the over-doped regime, in which the pairing strength becomes weak. Despite the highest Debye temperature of 400K is observed at $x = 0.4$, our theoretical volume bulk modulus of KFe$_2$As$_2$ is as weak as ~10GPa, causing a Debye temperature at 173K [21].

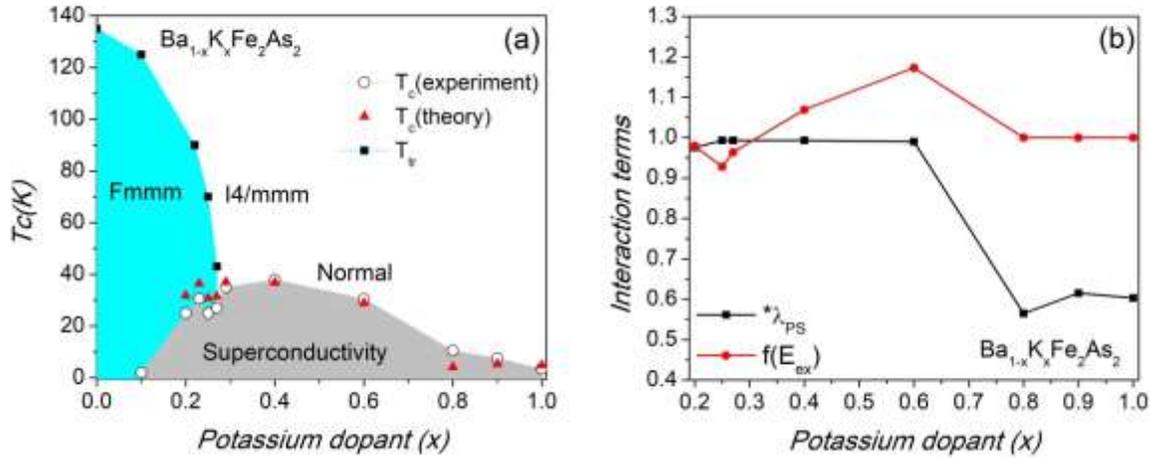

**Figure 3.** (a) The doping dependence of the theoretical and experimental [8,26] $T_c$ of Ba$_{1-x}$K$_x$Fe$_2$As$_2$. The $T_{tr}$ refer to the structural transition temperatures [26]. (b) The individual interaction terms of Ba$_{1-x}$K$_x$Fe$_2$As$_2$ as a function of K content. The Debye temperature in the limit of low temperature is used.

The $T_c$ calculation of the $Ba_{0.75}K_{0.25}Fe_2As_2$ in the $C_4$ magnetic phase is demonstrated below. The electron-phonon coupling at the Fermi surface $\lambda_{E_F}$, the Coh factor and the $R_g$ factor are 0.32, 4 and 5.9, respectively, and the complete list of $R_g$ factors can be found in the supplementary materials. The amplified electron-phonon coupling is $^*\lambda_{PS} = \frac{\lambda_{PS}}{\lambda_{PS}+1} = \frac{0.32*4^2*5.9^2}{0.32*4^2*5.9^2+1} = 0.994$ and the Coulomb pseudopotential is $\mu^* = \frac{\mu}{1+\lambda_{PS}} = \frac{0.15}{1+0.32*4^2*5.9^2} = 0.00084$. Doping is also associated with an internal chemical pressure and causes a variation of the antiferromagnetic exchange by $f(E_{ex}) \sim \frac{[M_{Fe}^2 E_{co}]_{P>0}}{[M_{Fe}^2 E_{co}]_{P=0}} = \left(\frac{1.48^2}{1.41^2}\right) 1.027 = 1.132$ in which the Fe moments are scaled to the Bohr magneton. Outside the $C_4$ magnetic phase range, the pairing strength is simply $\lambda = {}^*\lambda_{PS} f(E_{ex}) = 0.994*1.132 = 1.125$. However, in the $C_4$ magnetic phase in $Ba_{0.75}K_{0.25}Fe_2As_2$, the corrected pairing strength is given by $\lambda_{corrected} \sim \lambda \frac{E_{AF}(a)}{E_{AF}^{trial}(a)} = 1.125*0.82 = 0.922$ where $\frac{E_{AF}(c)}{E_{AF}^{trial}(c)}$ is neglected since $c \gg a$. We substitute the corrected pairing strength into the McMillian $T_c$ formula [31],

$$T_c = \frac{T_D}{1.45}\exp\left(\frac{-1.04(1+\lambda_{corrected})}{\lambda_{corrected}-\mu^*(1+0.62\lambda_{corrected})}\right) = \frac{393}{1.45}\exp\left(\frac{-1.04(1+0.922)}{0.922-(0.00084)(1+0.62(0.922))}\right) = 30.9K$$

**Discussion**

The increase of the amplified electron-phonon coupling in $BaFe_2As_2$ is mainly due to the $R_g$ factor. This is because the larger Debye frequency allows more energetic phonons to scatter more electrons below the Fermi surface [13]. However, the antiferromagnetic fluctuations are suppressed by high pressure, which compensates for the effect of $R_g$. This reduces the $T_c$ of $BaFe_2As_2$ above 2GPa. The $R_g$ factor takes into account the electron energies located down to ~30meV (~350K) below the Fermi level. This approach still satisfies the hyperbolic tangent shape of the Fermi-Dirac statistics across the Fermi level at $T > 0K$. The pairing strength formulae of $BaFe_2As_2$ (122-type), FeSe (11 type), NaFeAs (111-type) and LiFeAs (111-type) are identical [13], and therefore we believe that the pairing mechanism of these four undoped iron-based superconductors are the same. The experimental $T_c$ of $Ba_{0.8}K_{0.2}Fe_2As_2$ reported by Rotter *et al.* [26] and Böhmer *et al.* [8] are 26K and 20K, respectively. The small deviation of the experimental $T_c$ values may be related to the experimental methods used or due to the uncertainly in the tetrahedral Fe-As-Fe angle. If we use the tetrahedral angle of $BaFe_2As_2$ measured by R. Mittal *et al.* [36] at low temperatures as a reference material, the theoretical $T_c$ of $Ba_{0.8}K_{0.2}Fe_2As_2$ returns to ~18K. However, the use of Rotter *et al.* data provides a more systematic $T_c$ calculation since all lattice parameters, bond lengths and bond angles at $0 \leq x \leq 1$ are obtained from one experimental setup [26]. The dip-like structure in $T_c$ at $0.24 < x < 0.29$ in Figure 3a is due to the redistribution of local exchange coupling. When the orientation of the magnetic moments is switched from head-to-tail (HT) to side-by-side (SS) configuration along the *b*-axis, the coefficients of the energy transfer function as a function of space remains unchanged, since the magnetostriction of $Ba_{1-x}K_xFe_2As_2$ is small. Since the local antiferromagnetic exchange coupling at $x = 0.25$ is weaker than at $x = 0.27$, the $T_c$ of $Ba_{0.8}K_{0.25}Fe_2As_2$ is lower than that of $Ba_{0.8}K_{0.27}Fe_2As_2$. Although the theoretical $T_c$ values of $Ba_{1-x}K_xFe_2As_2$ are slightly overestimated in the low doping region in Figure 3a, the theoretical $T_c$ profile is reasonably consistent with the experimental data [8]. Our simulation of the $Ba_{0.6}K_{0.4}Fe_2As_2$ shows that the highest $\lambda_{E_F}$ is 0.8, which is comparable to other reported values [29]. The massive decrease in $^*\lambda_{PS}$ at $x > 0.8$ is due to the heavy fermion factor [19-21]. The Kondo temperature $T^*$ of the K-doped $BaFe_2As_2$ emerges when the degree of doping is higher than 0.73 [21]. In the presence of a heavy-fermion background, the local Fe moment at $T \sim 0K$ is much weaker than the electron spin [30]. As a result, the ultra-weak local Fe moment with the strength proportional to $1-\left(1-\frac{T}{T^*}\right)^{\frac{3}{2}}$ breaks the crosslink between the antiferromagnetism and electron-phonon coupling [20]. The consequence is that the electron-phonon coupling is amplified only by the out-of-plane phonon actuated by the tetrahedral site with $C_F = 2$. In the absence of a significant amount of antiferromagnetic energy, electron excitations well below the Fermi level are rare and therefore the $R_g$ factor may not be appropriate in the heavy-fermion situation.

A discrepancy between the experimental and theoretical $T_c$ can be observed at $x = 0.8$, which is due to the fact that our simulation does not take into account the increase in the Kondo temperature as a function of doping. We assume that the Kondo temperature is sufficiently high so that the local Fe moment is well screened. However, the measured Kondo temperature of $Ba_{0.27}K_{0.73}Fe_2As_2$ is only ~10K [21], so that the Kondo screening at $T_c$ may not be complete. The incomplete decoupling between antiferromagnetism and electron-phonon coupling therefore leads to a small underestimation of the $T_c$ of $Ba_{0.2}K_{0.8}Fe_2As_2$. This is not the case with the end member $KFe_2As_2$ for where the Kondo temperature is up to ~150K [19]. Then the residual Fe moments are screened much more effectively and thus the experimental and theoretical $T_c$ values in the doping range $0.8 < x < 1$ correspond better.

It is still disputed whether $KFe_2As_2$ is an *s*-wave or *d*-wave superconductor [37,38]. Considering the *s*-wave pairing in our model, the theoretical $T_c$ is 4.8K, which is the value included in Figure 3. We have also calculated the $T_c$ of $KFe_2As_2$ under *d*-wave order parameter symmetry. If for example $d_{x^2-y^2}$ pairing is considered, the electron phonon coupling at the Fermi energy is reduced from 0.38 to 0.27 [39]. After actuating the pairing potential through the tetrahedral sites, the pairing strength becomes 1.08 and the theoretical $T_c$ reaches 2.8K. Both values are relatively close to the experimental value, so that no clear statement about the pairing symmetry can be made from our model.

**Conclusion**

Based on ab-initio calculated parameters, we presented a theoretical model for the calculation of the superconducting transition temperatures of $BaFe_2As_2$ as a function of pressure and of $Ba_{1-x}K_xFe_2As_2$ as a function of K content $x$, which correspond reasonably well to the experimental values. Our model is able to account for details in their phase diagrams, e.g. the $T_c$ reduction by the presence of the $C_4$-reentrant tetragonal phase in the under-doped region or by the electronic mass enhancement and Kondo effect in the heavily over-doped region. It considers that when the electrons at the Fermi energy interact with the local magnetic moments of Fe, the spin density wave and the abnormal soft out-of-plane lattice vibrations cause a four-fold increase in the electron-phonon coupling.

**References**

[1] S. Sharma, S. Bharathia, S. Chandra, V. R. Reddy, S. Paulraj, A. T. Satya, V. S. Sastry, A. Gupta, C. S. Sundar, *Phys. Rev. B* **81**, 174512 (2010).

[2] S. Kasahara, T. Shibauchi, K. Hashimoto, K. Ikada, S. Tonegawa, R. Okazaki, H. Shishido, H. Ikeda, H. Takeya, K. Hirata, T. Terashima, Y. Matsuda, *Phys. Rev. B* **81**, 184519 (2010).

[3] M. Rotter, M. Tegel, D. Johrendt, *Phys. Rev. Lett.* **101**, 107006 (2008).

[4] A. S. Sefat, R. Jin, M. A. McGuire, B. C. Sales, D. J. Singh, D. Mandrus, *Phys. Rev. Lett.* **101**, 117004 (2008).

[5] J. Paglione, R. L. Greene, High-temperature superconductivity in iron-based materials, *Nat. Phys.* **6**, 645 – 658 (2010).

[6] J. M. Allred, K. M. Taddei, D. E. Bugaris, M. J. Krogstad, S. H. Lapidus, D. Y. Chung, H. Claus, M. G. Kanatzidis, D. E. Brown, J. Kang, R. M. Fernandes, I. Eremin, S. Rosenkranz, O. Chmaissem, R. Osborn, Double-Q spin-density wave in iron arsenide superconductors, *Nat. Phys.* **12**, 493–498 (2016).

[7] J. Hou, C.-w. Cho, J. Shen, P. M. Tam, I-H. Kao, M. H. G. Lee, P. Adelmann, T. Wolf, R. Lortz, Possible coexistence of double-Q magnetic order and chequerboard charge order in the re-entrant tetragonal phase of $Ba_{0.76}K_{0.24}Fe_2As_2$, *Physica C* **539**, 30–34 (2017).

[8] A. E. Böhmer, F. Hardy, L. Wang, T. Wolf, P. Schweiss, C. Meingast, Superconductivity-induced re-entrance of the orthorhombic distortion in $Ba_{1-x}K_xFe_2As_2$, *Nat. Commun.* **6**:7911(2015).

[9] A. E. Böhmer, A. Kreisel, Nematicity, magnetism and superconductivity in FeSe, *J. Phys.: Condens. Mat.* **30**, 023001 (2018).

[10] S. Deng, J. Köhler, A. Simon, Electronic structure and lattice dynamics of NaFeAs, *Phys. Rev. B* **80**, 214508 (2009).

[11] B. Li, Z. W. Xing, G. Q. Huang, M. Liu, Magnetic-enhanced electron-phonon coupling and vacancy effect in "111"-type iron pnictides from first-principle calculations, *J. App. Phys.* **111**, 033922 (2012).

[12] S. Coh, M. L. Cohen, S. G. Louie, Anti-ferromagnetism enables electron-phonon coupling in iron-based superconductors, *Phys. Rev. B* **94**, 104505 (2016).

[13] C. H. Wong, R. Lortz, The antiferromagnetic and phonon-mediated model of the NaFeAs, LiFeAs and FeSe superconductors, arXiv:1902.06463 (2019).

[14] X.-W. Jia, H.-Y. Liu, W.-T. Zhang, L. Zhao *et al.*, Common Features in Electronic Structure of the Oxypnictide Superconductors from Photoemission Spectroscopy, *Chin. Phys. Lett.* **25**, 3765-3768 (2008).

[15] C. Liu, G. D. Samolyuk, Y. Lee, N. Ni, T. Kondo, A. F. Santander-Syro, S. L. Bud'ko, J. L. McChesney, E. Rotenberg, T. Valla, A. V. Fedorov, P. C. Canfield, B. N. Harmon, A. Kaminski, K-doping dependence of the Fermi surface of the iron-arsenic $Ba_{1-x}K_xFe_2As_2$ superconductor using angle-resolved photoemission spectroscopy, *Phys Rev Lett.* **101**, 17 (2008)

[16] C. Zhang *et al.*, Ubiquitous strong electron–phonon coupling at the interface of $FeSe/SrTiO_3$, *Nat. Commun.* **8**, 14468 (2017).

[17] U. Stockert, M. Abdel-Hafiez, D. V. Evtushinsky, V. B. Zabolotnyy, A. U. B. Wolter, S. Wurmehl, I. Morozov, R. Klingeler, S. V. Borisenko, B. Büchner, Specific heat and angle-resolved photoemission spectroscopy study of the superconducting gaps in LiFeAs, *Phys. Rev. B* **83**, 224512 (2011).

[18] A. Mani, N. Ghosh, S. Paulraj, A. Bharathi, C. S. Sundar, Pressure-induced superconductivity in $BaFe_2As_2$ single crystal, *Europhys. Lett.* **87**, 17004 (2009).

[19] F. Hardy, A. E. Böhmer, D. Aoki, P. Burger, T. Wolf, P. Schweiss, R. Heid, P. Adelmann, Y. X. Yao, G. Kotliar, J. Schmalian and C. Meingast, Evidence of Strong Correlations and Coherence-Incoherence Crossover in the Iron Pnictide Superconductor $KFe_2As_2$, *Phys. Rev. Lett* **111**, 027002 (2013).

[20] Y.-f. Yang, D. Pines, Emergent states in heavy-electron materials, *PNAS* **109**, E3060 – E3066 (2012).

[21] V. Grinenko, P. Materne, R. Sarkar, H. Luetkens, K. Kihou, C. H. Lee, S. Akhmadaliev, D. V. Efremov, S.-L. Drechsler, H.-H. Klauss, Superconductivity with broken time-reversal symmetry in ion-irradiated $Ba_{0.27}K_{0.73}Fe_2As_2$ single crystals, *Phys. Rev. B* **95**, 214511 (2017).

[22] H. Takahashi, K. Igawa, K. Arii, Y. Kamihara, M. Hirano, H. Hosono, Superconductivity at 43 K in an iron-based layered compound $LaO_{12x}F_xFeAs$, *Nature (London)* **453**, 376-378 (2008).

[23] M. R. Ebrahimi and H. Khosroabadi, Effects of Fluorine Doping and Pressure on the Electronic Structure of $LaO_{1-x}F_xFeAs$ Superconductor: a First Principle Study, *J. Supercond. Nov. Magn.* **30**, 2065-2071 (2017).

[24] H. Takahashi, T. Tomita, H. Takahashi, Y. Mizuguchi, Y. Takano, S. Nakano, K. Matsubayashi, Y. Uwatoko, High-pressure studies on $T_c$ and crystal structure of iron chalcogenide superconductors, *Scie. Techn. Adv. Mat.* **13**, 054401 (2012).

[25] A F Wang, Z J Xiang, J J Ying, Y J Yan, P Cheng, G J Ye, X G Luo and X H Chen, Pressure effects on the superconducting properties of single-crystalline Co doped NaFeAs, *New J. Phys.* **14**, 113043 (2012).

[26] M. Rotter, M. Panger, M. Tegel, D. Johrendt, Superconductivity and Crystal Structures of $(Ba_{1-x}K_x)Fe_2As_2$ (x = 0 - 1), *Angew. Chem. Int. Ed.* **47**, 7949 –7952 (2008).

[27] J. Richard Christman, Fundamentals of Solid States Physics, Wiley (1986).

[28] P. Blaha, *et al.*, WIEN2k, An Augmented Plane Wave + Local Orbitals Program for Calculating Crystal Properties (Karlheinz Schwarz, Techn. Universität Wien, Austria) (2018).

[29] J. P. Perdew, *et al.*, Atoms, molecules, solids, and surfaces: Applications of the generalized gradient approximation for exchange and correlation, *Phys. Rev. B* **46**, 6671 (1992).

[30] A. D. Becke, Density-functional exchange-energy approximation with correct asymptotic behavior, *Phys. Rev. A* **38**, 3098 (1988).

[31] W. L. McMillian, Transition Temperature of Strong-Coupled Superconductors, *Phys. Rev.* **167**, 331 (1968).

[32] E. J. König, P. Coleman, The Coulomb problem in iron based superconductors, arXiv:1802.10580 (2018).

[33] Y. Wen, D. Wu, Re. Cao, L. Liu, L. Song, The Third-Order Elastic Moduli and Debye Temperature of $SrFe_2As_2$ and $BaFe_2As_2$: a First-Principles Study, J. *Supercond. Nov. Magn.* **30**, 1749–1756 (2017).

[34] H. Oh, S. Coh, M. L. Cohen, Calculation of the specific heat of optimally K-doped $BaFe_2As_2$, arXiv:1502.00055v2 (2018).

[35] G. P. Malik, Superconductivity: a new approach based on the Bethe-Salpeter equation in mean field approach, *Series on Directions in Condensed Matter Physics* **21**, 55-56 (2016).

[36] R. Mittal, S. K. Mishra, S. L. Chaplot, S. V. Ovsyannikov, E. Greenberg, D. M. Trots, L. Dubrovinsky, Y. Su, Th. Brueckel, S. Matsuishi, H. Hosono, G. Garbarino, Ambient- and low-temperature synchrotron x-ray diffraction study of $BaFe_2As_2$ and $CaFe_2As_2$ at high pressures up to 56 GPa, *Phys. Rev. B* **83**, 054503 (2011).

[37] J.-Ph. Reid, M. A. Tanatar, A. Juneau-Fecteau, R. T. Gordon, S. René de Cotret, N. Doiron-Leyraud, T. Saito, H. Fukazawa, Y. Kohori, K. Kihou, C. H. Lee, A. Iyo, H. Eisaki, R. Prozorov, L. Taillefer, Universal Heat Conduction in the Iron Arsenide Superconductor $KFe_2As_2$: Evidence of a d-Wave State, *Phys. Rev. Lett.* **109**, 087001 (2012).

[38] F. F. Tafti, A. Juneau-Fecteau, M.-È. Delage, S. René de Cotret, J.-Ph. Reid, A. F. Wang, X.-G. Luo, X. H. Chen, N. Doiron-Leyraud, L. Taillefer, Sudden reversal in the pressure dependence of $T_c$ in the iron-based superconductor $KFe_2As_2$, *Nat. Phys.* **9**, 349–352 (2013).

[39] J. Song, J. F. Annett, Electron-phonon coupling and d-wave superconductivity in the cuprates, *Phys. Rev. B* **51**, 3840 (1995).

# Supplementary Materials

**Table 1:** The DFT parameters of $Ba_{1-x}K_xFe_2As_2$. Debye temperatures at 0K are listed.

| $x$ | $a$ (Å) | $c$ (Å) | Debye temp. (K) | Fe-As-Fe (deg) | $R_g$ | $\lambda_{EF}$ | $C_F$ |
|---|---|---|---|---|---|---|---|
| 0 | 3.963 | 13.022 | 379 | 111.2 | - | - | - |
| 0.2 | 3.938 | 13.186 | 382 | 110.8 | 3.002 | 0.25 | 4 |
| 0.25 | 3.931 | 13.223 | 393 | 109.8 | 5.901 | 0.32 | 4 |
| 0.27 | 3.928 | 13.221 | 394 | 109.7 | 5.901 | 0.33 | 4 |
| 0.4 | 3.909 | 13.212 | 400 | 109.3 | 5.904 | 0.84 | 4 |
| 0.6 | 3.889 | 13.512 | 292 | 107.7 | 5.536 | 0.21 | 4 |
| 0.8 | 3.858 | 13.720 | 184 | 107.3 | - | 0.12 | 2 |
| 0.9 | 3.852 | 13.760 | 180 | 106.9 | - | 0.37 | 2 |
| 1 | 3.840 | 13.840 | 173 | 106.7 | - | 0.38 | 2 |

**Table 2:** The DFT parameters of $BaFe_2As_2$. Debye temperatures at room temperature are listed.

| $P$(GPa) | $a$ (Å) | $b$ (Å) | $c$ (Å) | Debye temp. (K) | Fe-As-Fe (deg) | $R_g$ | $\lambda_{EF}$ | $C_F$ |
|---|---|---|---|---|---|---|---|---|
| 0 | 3.963 | 3.963 | 13.022 | 470 | 109.27 | N/A | - | - |
| 0.8 | 3.936 | 3.936 | 12.916 | 474 | 109.27 | 2.006 | 0.33 | 4 |
| 1.3 | 3.921 | 3.943 | 12.852 | 477 | 109.25, 108.95 | 1.978 | 0.37 | 4 |
| 2 | 3.906 | 3.930 | 12.809 | 494 | 108.96, 108.63 | 4.903 | 0.33 | 4 |
| 3 | 3.895 | 3.923 | 12.738 | 504 | 108.92, 108.53 | 4.002 | 0.26 | 4 |
| 4 | 3.884 | 3.912 | 12.682 | 518 | 108.86, 108.47 | 5.444 | 0.30 | 4 |
| 5.5 | 3.865 | 3.894 | 12.600 | 534 | 108.81, 108.41 | 9.850 | 0.36 | 4 |
| 7.2 | 3.846 | 3.876 | 12.5241 | 544 | 108.79, 108.37 | 13.111 | 0.19 | 4 |